\documentclass[12pt,a4paper,twoside]{article}
\usepackage{graphics,references}
\footnotesep=\baselineskip
\date{\thanks{Astrophysical Dynamics 1999/2000,
              Alessandro B. Romeo (Ed.),
              Onsala Space Observatory,
              2000.}}
\begin{document}

\title{Workshop Presentation and Hot Topics}

\author{Alessandro B. Romeo\\[.33cm]
        Onsala Space Observatory\\
        Chalmers University of Technology\\
        SE-43992 Onsala, Sweden\\
        (romeo@oso.chalmers.se)}

\pagestyle{myheadings}

\markboth{Workshop Presentation and Hot Topics}
         {Alessandro B. Romeo}

\maketitle

\begin{abstract}

   The workshop `Astrophysical Dynamics 1999/2000' followed a homonymous
advanced research course, and both activities were organized by me.  In this
opening paper of the proceedings book, I describe them and document their
\emph{strong} impact on the academic life of the local institutions.

\end{abstract}

\section{Introduction}

The advanced research course was held at Chalmers University of Technology
and G\"{o}te\-borg University in October--December 1999; it was open to
graduate students, senior researchers, and motivated under-graduate students
with good background in physics and mathematics.  The course covered several
multi-disciplinary issues of modern research on astrophysical dynamics, and
thus also of interest to physicists, mathematicians and engineers.  The major
topic was gas dynamics, viewed in context with stellar dynamics and plasma
physics.  The course was complemented by parallel seminars on hot topics
given by experts in such fields, and open to a wide scientific audience.  In
particular, I gave a friendly introduction to \emph{wavelets}, which are
becoming an increasingly powerful tool not only for processing signals and
images but also for analysing fractals and turbulence, and which promise to
have important applications to dynamical modelling of disc galaxies.  The
course is presented in more detail in Sect.\ 2.  The basic reference is Romeo
(1999b); see also the references cited therein and in Sect.\ 2, and Romeo
(1999a) for a discussion of my ideas about teaching (in a different context).

   The workshop was held at Onsala Space Observatory on 12 and 13 January,
and 1 March, 2000; it was open to a wide scientific audience.  The workshop
with published proceedings book was, as a matter of fact, the
\emph{innovative} form of exam that I proposed for the advanced research
course.  The contributions were refereed and their quality is high on
average, exceptionally high in a few cases.  The workshop is presented in
more detail in Sect.\ 3.

   The advanced research course and the workshop all together produced great
enthusiasm in the students and welcomed the participation of a hundred
different people, which means \emph{an order of magnitude} more than an
average graduate course at Chalmers University of Technology and G\"{o}teborg
University.  What else should I say?  Enjoy reading the proceedings book!

\section{The Advanced Research Course}

\subsection{Lectures}

\begin{enumerate}
\item Basics about fluids (2 hours).
\item The equations of motion (4 hours).
\item Simple applications (4 hours).
\item Instabilities (4 hours).
\item Turbulences and fractals (2 hours).
\item Astrophysical fractals: interstellar medium and galaxies (1 hour).
      \begin{itemize}
      \item References: Combes (1999b).
      \end{itemize}
\item Which thermal physics for gravitationally unstable media? (1 hour).
      \begin{itemize}
      \item References: Pfenniger (1998).
      \end{itemize}
\item Shocks (4 hours).
\item Magnetic fields (4 hours).
\item Gas dynamics vs.\ stellar dynamics and plasma physics (2 hours).
\end{enumerate}
References: Shu (1992), unless otherwise specified; see also van Dyke (1982),
Shore (1992), Dyson \& Williams (1997) and Choudhuri (1998).

\subsection{Exercises}

\begin{enumerate}
\item Birth, life (and death?) of a galaxy: a step-by-step problem (6
      hours).
\end{enumerate}
References: Padmanabhan (1996); see also Binney \& Tremaine (1987) and Binney
\& Merrifield (1998).

\subsection{Parallel Seminars on Hot Topics}

\begin{enumerate}
\item `Wavelets: A Presentation for Scientists' -- Alessandro Romeo (2
      hours).
      \begin{itemize}
      \item Multimedia: computer-projector show using the Matlab Wavelet
            Toolbox, running on a high-performance Linux machine; public
            computer lab.
      \item References: Press et al.\ (1992), Vetterling et al.\ (1992),
            Misiti et al.\ (1997), Hubbard (1998), Mallat (1998), Bergh et
            al.\ (1999) and Wavelet Digest.
      \end{itemize}
\item `Black-Hole Accretion Discs' -- Marek Abramowicz (2 hours).
\item `Wavelets at Work in Physics' -- Alessandro Romeo (2 hours).
      \begin{itemize}
      \item Multimedia: computer-projector show using the Matlab Wavelet
            Toolbox, running on a high-performance Linux machine; public
            computer lab.
      \item References: Bowman \& Newell (1998), Fang \& Thews (1998),
            Goedecker (1998) and van den Berg (1999).
      \end{itemize}
\item `Small-Scale Structure and Dynamics in the Interstellar Medium' -- John
      Black (2 hours).
\item `Wavelets at Work in Astrophysics' -- Alessandro Romeo (2 hours).
      \begin{itemize}
      \item Photographs:
            \verb=http://www.mvd.chalmers.se/~pergus/alessandro/= .
      \item Multimedia: computer-projector show using the Matlab Wavelet
            Toolbox, running on a high-performance Linux machine, and the
            documentation of the MR/1-MR/2 Software Packages; public computer
            lab.
      \item References: Fang \& Thews (1998), Starck et al.\ (1998) and van
            den Berg (1999).
      \end{itemize}
\item `Magnetic Fields in Galaxies' -- Cathy Horellou (2 hours).
\item `Dynamical Modelling of Disc Galaxies: Multi-Scale Structures' --
      Alessandro Romeo (2 hours).
      \begin{itemize}
      \item References: Romeo (1994), Friedli (1996), Zhang (1996), Masset \&
            Tagger (1997), Romeo (1997, 1998a, b), Zhang (1998), Block \&
            Puerari (1999), Combes (1999a), Erwin \& Sparke (1999), Friedli
            (1999), Martini \& Pogge (1999), Shlosman (1999), Zhang (1999),
            Combes (2000), Englmaier \& Shlosman (2000), Maciejewski \&
            Sparke (2000) and Shlosman (2000).
      \end{itemize}
\end{enumerate}

\section{The Workshop}

\begin{enumerate}
\item `Turbulence and Fractal Analysis Using Wavelets' -- Michael F\"{o}rsth
      (1.5 hours).
      \begin{itemize}
      \item Proceedings: review paper.
      \item Referees: J\"{o}ran Bergh, Cathy Horellou and Alessandro Romeo.
      \end{itemize}
\item `Dark Matter and Cold Fractal Clouds' -- Achim Tappe (1.5 hours).
      \begin{itemize}
      \item Proceedings: review paper.
      \item Referees: John Black, Alessandro Romeo and Tommy Wiklind.
      \end{itemize}
\item `Fractals and Large-Scale Structure of the Universe' -- \v{G}irts
      Barinovs (1.5 hours).
      \begin{itemize}
      \item Proceedings: paper.
      \item Referees: Alessandro Romeo and Tommy Wiklind.
      \end{itemize}
\item `Magnetohydrodynamic Turbulence in Accretion Discs' -- Rim Turkmani
      (1.5 hours).
      \begin{itemize}
      \item Proceedings: review paper.
      \item Referees: Arto Heikkil\"{a}, Alessandro Romeo and Ulf Torkelsson.
      \end{itemize}
\item `Galaxy Collisions' -- Nils Tarras-Wahlberg (1.5 hours).
      \begin{itemize}
      \item Proceedings: abstract.
      \item Referees: Alessandro Romeo.
      \end{itemize}
\item `The Outflow-Disc Interaction in Young Stellar Objects' -- Michele
      Pestalozzi (1.5 hours).
      \begin{itemize}
      \item Proceedings: review paper.
      \item Referees: John Conway, Michael Olberg and Alessandro Romeo.
      \end{itemize}
\item `Jets from Herbig-Haro Objects' -- Jiyune Yi (1.5 hours).
      \begin{itemize}
      \item Proceedings: abstract.
      \item Referees: Alessandro Romeo.
      \end{itemize}
\item `Origin and Propagation of Extragalactic Jets' -- Alessandro Laudati
      (1.5 hours).
      \begin{itemize}
      \item Proceedings: abstract.
      \item Referees: Alessandro Romeo.
      \end{itemize}
\item `Why Are Evolved-Stellar Atmospheres Clumped?' -- Liz Humphreys
      (cancelled).
\item `Ionization Fronts and Photo-Dissociation' -- Henrik Olofsson (1.5
      hours).
      \begin{itemize}
      \item Proceedings: review paper.
      \item Referees: John Black, {\AA}ke Hjalmarson and Alessandro Romeo.
      \end{itemize}
\end{enumerate}

\section*{Acknowledgements}

I would very much like to thank the following people for their invaluable
help: the speakers of the parallel seminars on hot topics, the contributors
to the workshop and the referees of the proceedings; Christer Andersson
(guided visits), Per Gustafson (IT-project), Bert Hansson (transports),
H{\aa}kan H{\aa}kansson (AV-service), Bi\"{o}rn Nilsson and Michael Olberg
(computers), Michele Pestalozzi and Achim Tappe (support from A to Z!).  Last
but not least, I acknowledge the financial support of the Swedish Natural
Science Research Council, of the local institutions and the further generous
grants by the `Solveig och Karl G Eliassons Minnesfond'.


\begin{references}

\reference Bergh J., Ekstedt F., Lindberg M., 1999,
           Wavelets.  Studentlitteratur, Lund

\reference Binney J., Merrifield M., 1998,
           Galactic Astronomy.  Princeton University Press, Princeton

\reference Binney J., Tremaine S., 1987,
           Galactic Dynamics.  Princeton University Press, Princeton (Errata
           in arXiv:astro-ph/9304010)

\reference Block D.L., Puerari I., 1999,
           A\&A 342, 627

\reference Bowman C., Newell A.C., 1998,
           Rev.\ Mod.\ Phys.\ 70, 289

\reference Choudhuri A.R., 1998,
           The Physics of Fluids and Plasmas: An Introduction for
           Astrophysicists.  Cambridge University Press, Cambridge

\reference Combes F., 1999a,
           ASP Conf.\ Ser.\ 187, 59

\reference Combes F., 1999b,
           arXiv:astro-ph/9906477

\reference Combes F., 2000,
           ASP Conf.\ Ser.\ 197, 15

\reference Dyson J.E., Williams D.A., 1997,
           The Physics of the Interstellar Medium.  Institute of Physics
           Publishing, Bristol

\reference Englmaier P., Shlosman I., 2000,
           ApJ 528, 677

\reference Erwin P., Sparke L.S., 1999,
           ApJ 521, L37

\reference Fang L.-Z., Thews R.L.\ (Eds.), 1998,
           Wavelets in Physics.  World Scientific, Singapore

\reference Friedli D., 1996,
           ASP Conf.\ Ser.\ 91, 378

\reference Friedli D., 1999,
           ASP Conf.\ Ser.\ 187, 88

\reference Goedecker S., 1998,
           Wavelets and Their Application for the Solution of Partial
           Differential Equations in Physics.  Presses Polytechniques et
           Universitaires Romandes, Lausanne

\reference Hubbard B.B., 1998,
           The World According to Wavelets: The Story of a Mathematical
           Technique in the Making.  Peters, Wellesley

\reference Maciejewski W., Sparke L.S., 2000,
           MNRAS 313, 745

\reference Mallat S., 1998,
           A Wavelet Tour of Signal Processing.  Academic Press, San Diego

\reference Martini P., Pogge R.W., 1999,
           AJ 118, 2646

\reference Masset F., Tagger M., 1997,
           A\&A 322, 442

\reference Matlab Wavelet Toolbox:
           \verb=http://www.mathworks.com/products/wavelet/=

\reference Misiti M., Misiti Y., Oppenheim G., Poggi J.-M., 1997,
           Wavelet Toolbox for Use with Matlab: User's Guide.  The MathWorks,
           Natick

\reference MR/1-MR/2 Software Packages:
           \verb=http://www.multiresolution.com/=

\reference Padmanabhan T., 1996,
           Cosmology and Astrophysics through Problems.  Cambridge University
           Press, Cambridge

\reference Pfenniger D., 1998,
           Mem.\ S.A.It.\ 69, 429

\reference Press W.H., Teukolsky S.A., Vetterling W.T., Flannery B.P., 1992,
           Numerical Recipes in Fortran: The Art of Scientific Computing.
           Cambridge University Press, Cambridge

\reference Romeo A.B., 1994,
           A\&A 286, 799

\reference Romeo A.B., 1997,
           A\&A 324, 523

\reference Romeo A.B., 1998a,
           A\&A 335, 922

\reference Romeo A.B., 1998b,
           $N$-Body Simulations of Disc Galaxies Can Shed Light on the
           Dark-Matter Problem.  In: Salucci P.\ (ed.) Dark Matter.  Studio
           Editoriale Fiorentino, Firenze, p.\ 177

\reference Romeo A.B., 1999a,
           arXiv:physics/9906028

\reference Romeo A.B., 1999b,
           Astrophysical Dynamics.  Course Notes, Onsala Space Observatory
           (unpublished)

\reference Shlosman I., 1999,
           ASP Conf.\ Ser.\ 187, 100

\reference Shlosman I., 2000,
           ASP Conf.\ Ser.\ 197, 23

\reference Shore S.N., 1992,
           An Introduction to Astrophysical Hydrodynamics.  Academic Press,
           San Diego

\reference Shu F.H., 1992,
           The Physics of Astrophysics -- Vol.\ II: Gas Dynamics.  University
           Science Books, Sausalito

\reference Starck J.-L., Murtagh F., Bijaoui A., 1998,
           Image Processing and Data Analysis: The Multiscale Approach.
           Cambridge University Press, Cambridge

\reference van den Berg J.C.\ (Ed.), 1999,
           Wavelets in Physics.  Cambridge University Press, Cambridge

\reference van Dyke M., 1982,
           An Album of Fluid Motion.  Parabolic Press, Stanford

\reference Vetterling W.T., Teukolsky S.A., Press W.H., Flannery B.P., 1992,
           Numerical Recipes Example Book (Fortran).  Cambridge University
           Press, Cambridge

\reference Wavelet Digest: \verb=http://www.wavelet.org/=

\reference Zhang X., 1996,
           ApJ 457, 125

\reference Zhang X., 1998,
           ApJ 499, 93

\reference Zhang X., 1999,
           ApJ 518, 613

\end{references}
\end{document}